\documentclass{desyproc}

\begin{document}

\title{Axions and White Dwarfs}

\author{{\slshape J. Isern$^{1,2}$, 
                  S. Catal\'an$^3$, 
                  E. Garc\'{\i}a--Berro$^{4,2}$, 
                  M. Salaris$^5$, 
                  S. Torres$^{4,2}$ }\\[1ex]
          $^1$Institut de Ci\`encies de l'Espai (CSIC), 
              Campus UAB, 08193 Bellaterra, Spain \\
          $^2$Institut d'Estudis Espacials de Catalunya (IEEC), 
              Ed. Nexus, c/Gran Capit\`a, 08034 Barcelona, Spain\\
          $^3$Center for Astrophysics Research, University of Hertfordshire, 
              College Lane, Hatfield AL10 9AB, UK \\
          $^4$Departament de F\'{\i}sica Aplicada, 
              Universitat Polit\`ecnica de Catalunya,
              c/Esteve Terrades 5, 08860 Castelldefels, Spain\\
          $^5$Astrophysics Research Institute, 
              Liverpool John Moores University, 
              12 Quays House, Birkenhead, CH41 1LD, UK \\ }

\contribID{Isern\_Jordi}

\desyproc{DESY-PROC-2010-03}
\acronym{Patras 2010} 
\doi  

\maketitle

\begin{abstract}
White  dwarfs are  almost  completely degenerate  objects that  cannot
obtain energy  from the thermonuclear  sources and their  evolution is
just  a gravothermal  process  of cooling.   The  simplicity of  these
objects,  the fact that  the physical  inputs necessary  to understand
them are well identified, although not always well understood, and the
impressive observational  background about white dwarfs  make them the
most well studied Galactic population.  These characteristics allow to
use  them as  laboratories  to test  new  ideas of  physics.  In  this
contribution  we  discuss  the   robustness  of  the  method  and  its
application to the axion case.
\end{abstract}

\section{Introduction}

White  dwarfs are  almost  completely degenerate  objects that  cannot
obtain energy  from the thermonuclear  sources and their  evolution is
just a  gravothermal process of  cooling.  Globally, the  evolution of
the luminosity of a white dwarf can be written as:

\begin{equation}
L+L_{\nu} +L_{\rm X}=-  \int^{M_{\rm WD}}_0 C_{\rm v}\frac{dT}{dt}\,dm
- \int^{M_{\rm       WD}}_0      T\Big(\frac{\partial      P}{\partial
  T}\Big)_{V,X_0}\frac{dV}{dt}\,dm   +\;\;   (l_{\rm   s}+e_{\rm   s})
\dot{M}_{\rm s} + \dot{\epsilon}_{\rm X}
\end{equation}

\noindent where the  l.h.s.  of this equation represents  the sinks of
energy, photons,  neutrinos and any additional exotic  term, while the
r.h.s. contains the sources of  energy, the heat capacity of the star,
the work due  to the change of volume, the  contribution of the latent
heat  and   gravitationl  settling  upon   crystallization  (the  term
$\dot{M}_{\rm s}$  is the rate  of crystallization) and,  finally, the
last    term   represents    any   additional    exotic    source   of
energy~\cite{Isern:1998}.  There  are two ways  to test the  theory of
white  dwarf evolution,  namely, studying  the white  dwarf luminosity
function  and using  the  secular  drift of  the  pulsation period  of
variable white dwarfs.

During  the cooling process,  white dwarfs  experience some  phases of
pulsational   instability   powered   by   the   $\kappa$-   and   the
\emph{convective  driven}-mechanisms \cite{review}.  Depending  on the
composition of the atmosphere variable  white dwarfs are known as DOV,
DBV and DAV. These stars  also known as PG1159 or GW Vir
stars, V777 He  stars and ZZ Ceti stars  respectively.  Variable white
dwarfs  of  different compositions  occupy  different  regions in  the
Hertzsprung--Russell  diagram.   The  value  of the  pulsation  period
indicates  that  these objects  are  experiencing g--mode  non--radial
pulsations, where the main restoring force is gravity. One of the main
characteristics of these pulsations  is that they experience a secular
drift  caused by  the evolution  of the  temperature and  radius.  For
qualitative    purposes    this     drift    can    be    approximated
by~\cite{Winget:1983}:

\begin{equation}
\frac{d\ln \Pi}{dt} \simeq -a \frac{d\ln T}{dt} + b \frac{d\ln R}{dt}
\end{equation}

\noindent where $a$  and $b$ are constants of the  order of unity that
depend on  the details of the model,  and $R$ and $T$  are the stellar
radius  and  the  temperature  at  the  region  of  period  formation,
respectively.  This equation reflects the fact that, as the star cools
down, the degeneracy of  the plasma increases, the buoyancy decreases,
the   Brunt-V\"ais\"al\"a  frequency   becomes  smaller   and,   as  a
consequence,  the spectrum  of  pulsations gradually  shifts to  lower
frequencies. At  the same time,  since the star contracts,  the radius
decreases and the frequency tends to increase. In general, DAV and DBV
stars are  already so  cool (and degenerate)  that the radial  term is
negligible and the  change of the period of  pulsation can be directly
related  to  the change  in  the core  temperature  of  the star.  The
timescales involved are of the  order of $\sim 10^{-11}$ s/s for DOVs,
$\sim 10^{-13}$ to $\sim 10^{-14}$ s/s for DBVs and $\sim 10^{-15}$ to
$\sim  10^{-16}$ s/s  for  DAVs. Therefore,  the  measurement of  such
drifts  provides an  effective method  to test  the theory  of cooling
white dwarfs. This measurement is a difficult but feasible task, as
it  has been  already proved  in  the case  of G117--B15A,  a ZZ  Ceti
star~\cite{Kepler:2005}. These      properties      allow      to      build      a      simple
relationship~\cite{Isern:1992,Isern:2008} to estimate the influence of
an extra sink  of energy, axions for instance, on  the period drift of
variable white dwarfs: $
({L_{\rm X}}/{L_{\rm model}}) \approx ({\dot{\Pi}_{\rm obs}}/
{\dot{\Pi}_{\rm model}}) -1 
\label{eqf}
$
  where the  suffix ``model''  refers to  those  models built
using standard physics.

The white dwarf  luminosity function is defined as  the number density
of white dwarfs of a given luminosity per unit of magnitude interval:

\begin{equation}
n(l) = \int^{M_{\rm s}}_{M_{\rm i}}\,\Phi(M)\,\Psi(t)
\tau_{\rm cool}(l,M) \;dM
\label{lf}
\end{equation}

\noindent   where  $t$   satisfies   the  condition   $t  =   T-t_{\rm
  cool}(l,M)-t_{\rm PS}(M)$  and $l =  -\log (L/L_\odot)$, $M$  is the
mass of the parent star  (for convenience all white dwarfs are labeled
with the mass of the  main sequence progenitor), $t_{\rm cool}$ is the
cooling time down to luminosity $l$, $\tau_{\rm cool}=dt/dM_{\rm bol}$
is the  characteristic cooling time,  $M_{\rm s}$ and $M_{\rm  i}$ are
the maximum and the minimum masses  of the main sequence stars able to
produce a white dwarf of  luminosity $l$, $t_{\rm PS}$ is the lifetime
of  the progenitor  of the  white dwarf,  and $T$  is the  age  of the
population under  study.  The  remaining quantities, the  initial mass
function, $\Phi(M)$,  and the star formation rate,  $\Psi(t)$, are not
known a priori and depend  on the properties of the stellar population
under study. The computed luminosity function is usually normalized to
the bin with the smallest error bar, traditionally the one with $l=3$,
in  order to  compare theory  with observations.   Equation (\ref{lf})
shows  that in  order to  use the  luminosity function  as  a physical
laboratory it is necessary to  have good enough observational data and
to know  the galactic properties that  are used in  this equation (the
star  formation rate, the  initial mass  function and  the age  of the
Galaxy). Fortunately, the bright  branch of the luminosity function is
only  sensitive to the  average characteristic  cooling time  of white
dwarfs at  the corresponding luminosity when the  function is properly
normalized. The  reason is twofold~\cite{Isern:2009}, on  one hand the
stellar population is dominated by low mass stars and on the other the
lifetime of stars increases very  sharply when the mass decreases. The
result is that the old galactic populations are still producing bright
white dwarfs  and the number of  such stars at each  luminosity bin is
the sum  of the  contributions of all  the different episodes  of star
formation.

\begin{figure}[htb]
\centerline{\includegraphics[width=0.6\textwidth]{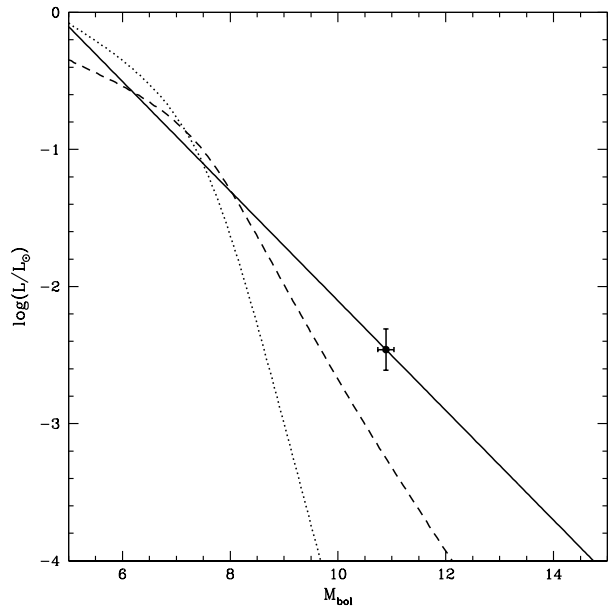}}
\caption{Power  emitted  in the  form  of  photons (continuous  line),
         neutrinos  (dotted line) and  axions (dashed  line, arbitrary
         $g_{\rm  aee}$) by  a 0.6  $M_\odot$ white  dwarf  versus its
         bolometric magnitude, a function that monotonically increases
         with   time.   The   cross   represents   the   position   of
         G115-B15A.}\label{Fig:lum}
\end{figure}

\section{The axion case}

We note  that in white  dwarf case only  DFSZ axions have to  be taken
into account, since electron bremsstrahlung is the dominant process in
white dwarf interiors.  The axion emission rate under these conditions
is given by~\cite{Nakagawa:1987}: $
\dot{\epsilon}_{\rm ax} = 1.08\times 10^{23}{(g^2_{\rm aee}}/{4\pi})
  ({Z^2}{A})\,T^4_7 F(T,\rho) \,\,\, {\rm erg/g/s}
$ 
where $F$ takes  into account  the Coulomb  plasma effects,
$T_7$ is  the temperature in  units of $10^7$  K, $Z$ and $A$  are the
atomic and  mass numbers of  the plasma components,  respectively, and
$g_{\rm aee}$  is the strength of the  axion-electron Yukawa coupling.
Figure \ref{Fig:lum} shows the energy  losses of a typical white dwarf
star. Since the neutrino  emission is $\dot{\epsilon}_\nu \propto T^8$
and $L_{\rm  phot} \propto T^{2.7}$ the luminosity  function allows to
disentangle  the contribution  of the  different mechanisms  of energy
loss. When this method is  applied to the luminosity function of white
dwarfs  the value  of  $g_{\rm  aee}$ that  best  fits the  luminosity
function is $1.1\times  10^{-13}$, but variations of a  factor two are
still                compatible                with                the
observations~\cite{Isern:2008a,Isern:2009}.  Finally, the  most recent
analysis of  G117-B15A shows  that the value  of $g_{\rm  aee}$ quoted
here is  compatible with  the secular drift  of the  pulsation period,
which gives support to the  necessity to include an extra cooling term
in the white dwarf models~\cite{Isern:2010}.
 
\section{Conclusions}

There are  two independent evidences, the luminosity  function and the
secular drift of  DAV white dwarfs, that these  stars are cooling down
more rapidly than predicted. The introduction of an additional sink of
energy  linked to  the interaction  of  electrons with  a light  boson
(axion,  ALP,\ldots)  with an  strength  $g_{\rm  aee} \sim  10^{-13}$
solves   the  problem   satisfactorily.  Of   course,   the  remaining
uncertainties,  both observational and  theoretical, still  prevent to
claim  the existence  of such  interaction. A  direct  detection under
laboratory  conditions,  like those  of  the  CAST  helioscope \cite{Papaev:2010} or  the
shining  through  the   wall  experiments \cite{Arias:2010},  could  provide
unambiguous  evidences.  In  this  sense,  if the  current  result  is
interpreted  as due  to axions,  such  particles should  have a  mass,
$m_{\rm a}  \sim {\rm  meV}$ and should  be coupled with  photons with
$g_{\rm a\gamma} \sim 10^{-12}$ GeV$^{-1}$.

\section*{Acknowledgments}

This work has  been supported by the MICINN  grants AYA08-1839/ESP and
AYA2008-04211-C02-01, by the ESF EUROCORES Program EuroGENESIS (MICINN
grant EUI2009-04170),  by SGR grants  of the Generalitat  de Catalunya
and by the EU-FEDER funds.

\begin{footnotesize}

\end{footnotesize}

\end{document}